\documentclass{article}

\usepackage{graphicx}

\usepackage[final, nonatbib]{neurips_2024}

\usepackage[utf8]{inputenc} % allow utf-8 input
\usepackage[T1]{fontenc}    % use 8-bit T1 fonts
\usepackage{hyperref}       % hyperlinks
\usepackage{url}            % simple URL typesetting
\usepackage{booktabs}       % professional-quality tables
\usepackage{amsfonts}       % blackboard math symbols
\usepackage{nicefrac}       % compact symbols for 1/2, etc.
\usepackage{microtype}      % microtypography
\usepackage{xcolor}         % colors

\title{Transfer Learning Adapts to Changing PSD in Gravitational Wave Data}

\author{%
  Beka Modrekiladze %\thanks{Use footnote for providing further information
    \\
  Carnegie Mellon University\\
  Pittsburgh, PA 15213 \\
  \texttt{bekam@cmu.edu} \\
}

\begin{document}

\maketitle

\begin{abstract}
The detection of gravitational waves has opened unparalleled opportunities for observing the universe, particularly through the study of black hole inspirals. These events serve as unique laboratories to explore the laws of physics under conditions of extreme energies. However, significant noise in gravitational wave (GW) data from observatories such as Advanced LIGO and Virgo poses major challenges in signal identification. Traditional noise suppression methods often fall short in fully addressing the non-Gaussian effects in the data, including the fluctuations in noise power spectral density (PSD) over short time intervals. These challenges have led to the exploration of an AI approach that, while overcoming previous obstacles, introduced its own challenges, such as scalability, reliability issues, and the vanishing gradient problem. Our approach addresses these issues through a simplified architecture. To compensate for the potential limitations of a simpler model, we have developed a novel training methodology that enables it to accurately detect gravitational waves amidst highly complex noise. Employing this strategy, our model achieves over 99\% accuracy in non-white noise scenarios and shows remarkable adaptability to changing noise PSD conditions. By leveraging the principles of transfer learning, our model quickly adapts to new noise profiles with just a few epochs of fine-tuning, facilitating real-time applications in dynamically changing noise environments.
\end{abstract}

\section{Introduction}

The discovery of gravitational waves (GWs) has revolutionized our approach to observing the universe, granting us access to phenomena that remain invisible to traditional astronomical instruments \cite{LIGOScientific:2016gtq}. Among these phenomena, black hole inspirals are particularly significant. They serve as cosmic laboratories where the fundamental laws of physics can be tested under conditions of extreme gravity. The insights gained from these events have the potential to deepen our understanding of the universe's most fundamental principles, including the nature of spacetime itself.

However, the path to unlocking these insights is fraught with challenges, chief among them being the significant noise that contaminates GW data \cite{Glanzer:2022avx}. This noise, a mixture of instrumental and environmental interferences \cite{Abbott:2016xvh}, often masks the very signals we seek to analyze, making it difficult to extract clear and reliable information from the data collected by observatories such as Advanced LIGO \cite{LIGOScientific:2014pky} and Virgo \cite{VIRGO:2014yos}.

Furthermore, the substantial volume of data necessitates real-time analysis; however, traditional methodologies frequently lag, struggling to maintain pace \cite{Usman:2015kfa}. In the context of signal detection, the presence of the non-stationary characteristics of noise \cite{LSC:2018vzm} poses a significant challenge, given that traditional matched filtering techniques are designed and optimized for Gaussian noise \cite{Jaranowski:2005hz}. Because noise in the data can drastically reduce the signal-to-noise ratio, this leads to challenges in accurately identifying genuine GW events \cite{Finn:1992wt}.

Traditional denoising methods in GW astronomy have predominantly centered on direct noise suppression techniques, incorporating a diverse array of approaches. These methods range from variational-based techniques \cite{Torres:2014zoa}, and wavelet-based strategies \cite{Cornish:2020dwh, Klimenko}, to those employing the Hilbert-Huang transform \cite{Akhshi:2020xmd}. Despite their effectiveness in specific contexts, these traditional approaches fall short in addressing the intricate and evolving nature of noise within GW data.

Previous studies introducing AI for enhancing GW signal detection have revealed promising avenues \cite{Huerta:2021ybd, Zhao:2023tqr}. Innovations such as denoising autoencoders \cite{Shen:2019ohi} have been utilized for reconstructing signals from noisy data, while Recurrent Neural Networks \cite{Chatterjee:2021lit} are employed to capture temporal dependencies. These algorithms are distinguished by their computational efficiency, utilizing accelerated hardware to achieve rapid solutions \cite{Benedetto:2023jwn}. Their scalability will ensure the capability to manage large datasets, thus offering reliable estimates of model performance \cite{Huerta:2020xyq}. Moreover, the modular design of these algorithms enhances their adaptability, enabling seamless integration of novel methodologies \cite{Ma:2022esx}. Additionally, the generalization abilities of deep learning models guarantee consistent performance across diverse gravitational wave data analysis scenarios \cite{McGinn:2021jqg}. 

However, while autoencoders demonstrate effectiveness, their reliance on unsupervised learning may introduce reliability issues for scientific purposes in the future. Similarly, Recursive Neural Networks are susceptible to the vanishing gradient problem \cite{Bengio1994} as the network depth increases, which poses challenges for realistic signal detection. These obstacles highlight the need for a solution that incorporates supervised learning and employs a simpler neural network architecture. Such an approach would facilitate scalability and circumvent the vanishing gradient problem, offering a more robust framework for gravitational wave signal detection.

In response to these challenges, this paper focuses on a straightforward multi-layer perceptron (MLP) model and employs the ReLU activation function, which is crucial for circumventing the vanishing gradient problem. However, with simpler models, perceiving gravitational waves amidst highly non-trivial noise becomes almost impossible. To address this, we introduce a novel training methodology. We first train the model on clean, noise-free data to establish a robust foundation. Then, harnessing the principles of transfer learning, we fine-tune the model on noisy data, starting from the pre-trained model on clean data. This novel training methodology allowed us to effectively detect gravitational wave signals amidst highly non-trivial noise, demonstrating that simplicity, coupled with innovative training techniques, can lead to impactful outcomes.

The paper starts with \nameref{bank}, detailing the creation of a diverse set of gravitational waveforms. This is followed by \textbf{AI Model Development and Training}, which includes detailed discussions on \nameref{dataprep}, \nameref{architecture} and \nameref{training}. Subsequently, the paper examines \textbf{Realistic Noise and Fine Tuning} where we simulate authentic observational conditions akin to those at LIGO detectors and adapt AI models through transfer learning to these realistic, variable noise conditions. \nameref{perf} assesses how effectively our AI models detect gravitational wave signals amidst noise, demonstrating their potential to transform gravitational wave astronomy. Finally, \nameref{next} advocates for the development of more extensive waveform banks and the use of generative models to surpass current limitations, potentially ushering in a new era of discovery-led science.

\section{Methodology}\label{methodology}
\subsection{Waveform generation}\label{bank}
Our project simulates binary black hole mergers to explore a wide range of gravitational waveforms. Utilizing the \texttt{pycbc} library \cite{Biwer:2018osg, pycbc}, we varied key astrophysical parameters—masses (\(m_1, m_2\)) from \(10M_{\odot}\) to \(30M_{\odot}\) and spins (\(s_1, s_2\)) from 0 (non-spinning) to 0.99 (near-maximal spin)—to produce plausible gravitational waveform morphologies, based on observed black hole binaries and theoretical predictions. \cite{LIGOScientific:2018mvr}.

We employed the SEOBNRv4 approximant \cite{Bohe:2016gbl} to generate accurate gravitational waveforms that represent the inspiral, merger, and ringdown phases. Waveforms were generated using the get\_td\_waveform function from the \texttt{pycbc.waveform} module at a sampling rate of 4096 Hz, starting from a 40 Hz lower frequency cutoff, aligning with the sensitivity range of Advanced LIGO.

\subsection{Model development and training}

\paragraph{Data preparation}\label{dataprep}
Our dataset comprised a collection of waveforms, each representing a variety of black hole inspiral events. To these, we added ’No Signal’ samples, effectively augmenting our dataset with examples where GWs are absent. To ensure all waveforms were of consistent length, zeros were padded to the end of the input waveforms after the inspiral phase. This standardized waveform length allows our model to operate effectively with real data, where the exact merger time may be unknown. By employing uniformly sized segments, the model can systematically inspect all template waveforms in a single run, sliding the same size segment to inspect for potential waveforms.

\paragraph{Model architecture}\label{architecture}
Our chosen architecture was a Multilayer Perceptron \cite{MLP}, celebrated for its proficiency in discerning patterns within high-dimensional data. The MLP featured an input layer to linearize the waveform data, followed by two hidden layers with ReLU activations \ref{table:simple_mlp_architecture}. To prevent overfitting and to encourage the model to learn more generalized patterns, dropout regularization \cite{Dropout} was integrated into the hidden layers. Note also that the size of the hidden dimension is significantly smaller than the input data, 
which forces the model to learn intricate patterns instead of just memorizing 
the waveform bank. This design decision aimed to bolster the model’s resilience when confronted with noisy data. 

\begin{table}
  \caption{Architecture of the SimpleMLP Model}
  \label{table:simple_mlp_architecture}
  \centering
  \begin{tabular}{lccr}
    \toprule
    \textbf{Layer (type)} & \textbf{Input Shape} & \textbf{Output Shape} & \textbf{Param \#} \\
    \midrule
    Flatten (flatten) & 4205 & 4205 & 0 \\
    Linear (fc1) & 4205 & 64 & 269,184 \\
    Dropout (dropout1) & 64 & 64 & 0 \\
    Linear (fc2) & 64 & 64 & 4,160 \\
    Dropout (dropout2) & 64 & 64 & 0 \\
    Linear (fc3) & 64 & 10 & 650 \\
    \midrule
    \textbf{Total Parameters} & & & \textbf{273,994} \\
    \bottomrule
  \end{tabular}
\end{table}

\paragraph{Training process}\label{training}

To further regularize our model and encourage the learning of more robust features, we introduced weight decay into the optimization process. \cite{Goodfellow} This served as a countermeasure against overfitting by penalizing large weights. Throughout numerous epochs, our model engaged in an iterative learning process, aimed at reducing the loss on the training set, all the while being benchmarked against the validation set to ensure reliable performance.

\paragraph{Evaluation on clean data}
Upon completion of the training, the model was subjected to a test dataset, where it demonstrated high accuracy in identifying and classifying GW signals correctly. Interestingly, models trained under 'harsher' conditions—such as with dropout regularization and smaller hidden dimensions—couldn't reach over 99\% accuracy on clean data. However, these models performed significantly better when exposed to noise they had not previously encountered, compared to models that achieved over 99\% accuracy on clean data. Thus, a training approach was selected that would intentionally aim for models to achieve "almost" perfect performance, optimizing their robustness in real-world noisy environments.

\subsection{Realistic noise and transfer learning}

\paragraph{Realistic noise}
To replicate the authentic observational conditions of LIGO detectors, we employed the aLIGOZeroDetHighPower Power Spectral Density model to generate realistic noise patterns \cite{pycbc, noise}. The PSD is a critical tool in delineating the noise profile of LIGO detectors, which details the power distribution over a range of frequencies. This model captures the multifaceted nature of detector noise, that stem from a myriad of sources such as thermal fluctuations, quantum uncertainties, and environmental disturbances. Each waveform in our dataset was paired with noise, which was generated using a unique random seed, mirroring the unpredictable and variable noise encountered in actual GW detection. This approach to noise simulation enhances the validity of our tests and challenges our AI models to perform under realistic conditions that include the full spectrum of disturbances—from seismic activity and thermal vibrations to instrumental artifacts. Such noise modeling is crucial in evaluating the resilience and accuracy of our AI models in discerning GW signals from the noisy background, which is typical of the data collected by GW observatories. 

\paragraph{Challenge: time varying PSD} 
In the real world, the ultimate operational domain for our models, PSD is subject to change over time. This means that even if we train our model on a specific PSD, its effectiveness could diminish the very next day, which explains the cautious integration of AI models. Intriguingly, this challenge not only highlights the complexity of our initial problem but also points us toward the elegant solution that simultaneously addresses the issue of varying PSD. This solutions could potentially enable the deployment of AI models that effectively detect signals in real-time amidst fluctuating noise conditions.

\paragraph{Solution: transfer learning and fine-tuning}
To enhance model performance in detecting GW signals amidst noise, we employed transfer learning and fine-tuning. Models were initially pre-trained on clean datasets to learn the fundamental patterns of gravitational waveforms, providing a solid foundation. We then used transfer learning to adapt these models to our specific task by fine-tuning them on datasets with realistic noise profiles. This fine-tuning adjusted model parameters to better detect subtle features in noisy data, similar to real-world GW detection scenarios. Starting with pre-trained weights rather than random ones allowed us to achieve sufficient accuracy in just a few epochs, significantly speeding up the process. This approach enables real-time detection of gravitational wave signals by fine-tuning the model with changing PSD.

\paragraph{Model performance and evaluation}\label{perf}
The final evaluation of our AI models was conducted on a dataset that reflected true operational challenges by incorporating realistic noise. Remarkably, the models demonstrated exceptional proficiency in signal detection, with over 99\% accuracy, even in the presence of complex noise that is known to impede traditional analysis methods like matched filtering. We illustrate a sample detection by our model, where a gravitational wave signal is accurately identified within a noisy environment. This level of performance not only attests to the effectiveness of our training approach but also suggests that AI could play a transformative role in gravitational wave astronomy, particularly in scenarios where matched filtering is less effective.

\begin{figure}[h]
   \centering
   % First image
   \includegraphics[width=0.7\textwidth]{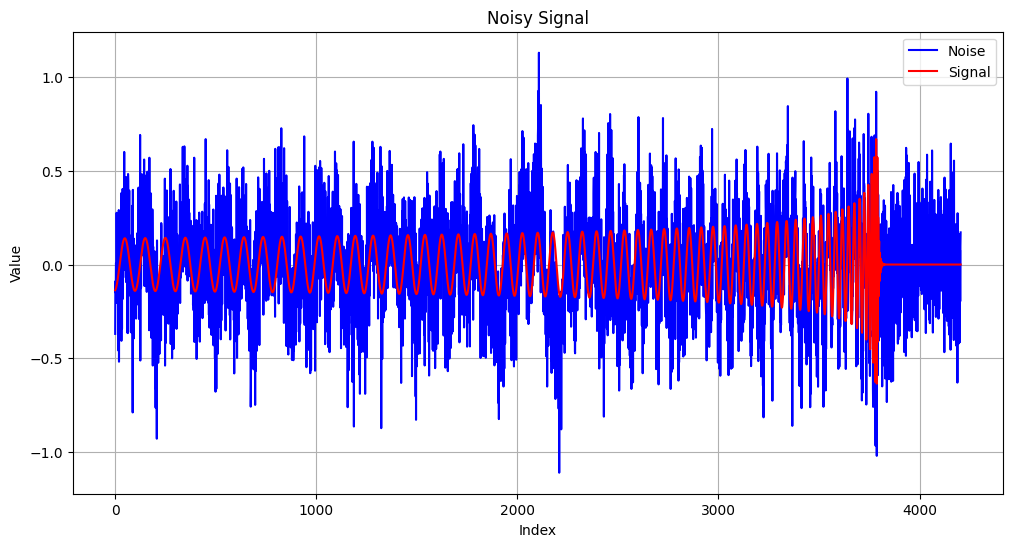}
   % Add vertical space between images
   % Second image
   \includegraphics[width=0.7\textwidth]{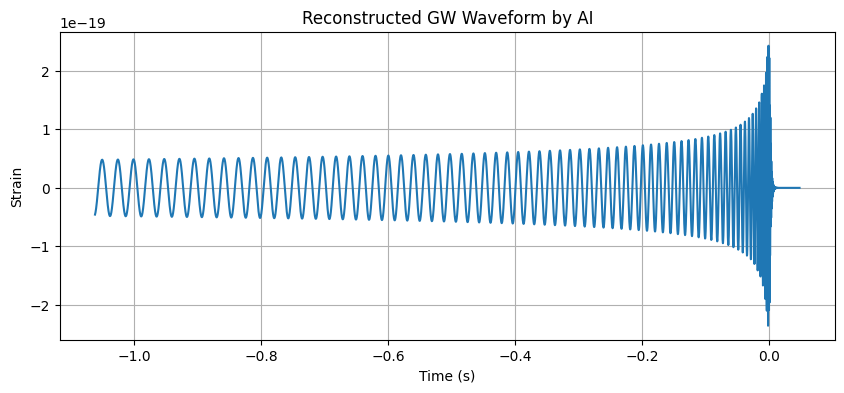}
   % Add vertical space between images
   % Third image
   \includegraphics[width=0.7\textwidth]{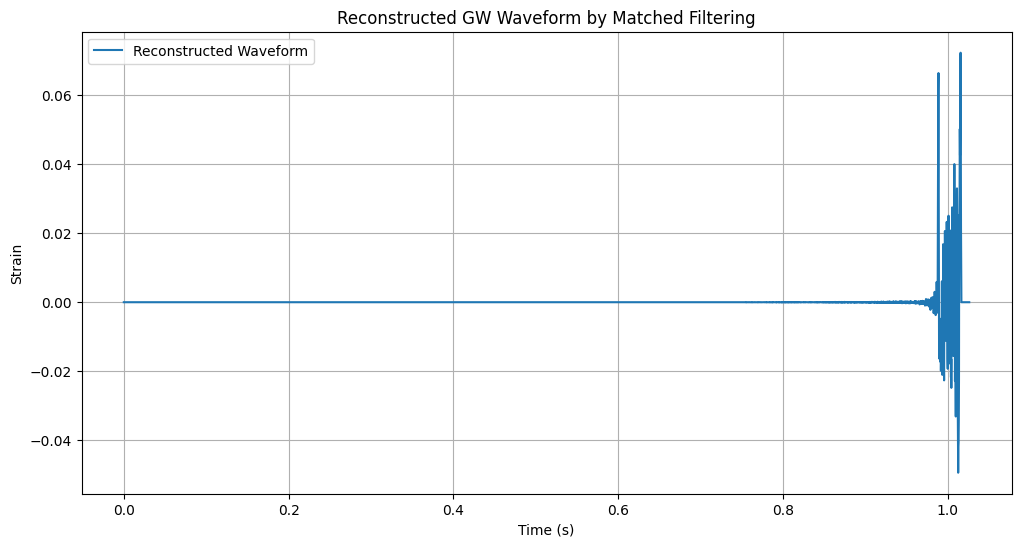}
   % Overall caption for all images
   \caption{Gravitational waveform detection in noisy signal.}
   \label{fig:combined_images}
\end{figure}

Figures below \ref{fig:combined_images} show the model's capability to discern and reconstruct the gravitational wave signal. The first figure shows a noisy data strain where our model detected a gravitational wave. The second and third figures display the reconstructed waveforms from the noisy data using our model and matched filtering, respectively, when the correct waveform is provided to matched filtering.

\section{Next steps and discussion }\label{next}

This success highlights AI's potential to reveal hidden signals in noise, opening new avenues for discoveries in GW astronomy. Moreover, this model can serve as a complement to matched filtering, rather than fully replacing it, facilitating a bidirectional enhancement of each other's capabilities. By assessing the probability of events in noisy data, our model identifies high-probability candidates, allowing matched filtering to conduct more focused and thorough investigations within shorter segments. Conversely, starting with an algorithmic denoising process and potentially exploring additional techniques, such as the method developed to correct, at first order, the effects of PSD variation on the search background \cite{Zackay:2019kkv}, prepares the data for more effective waveform detection by our model.

However, the current approach to GWs detection, which relies on waveform banks for matched filtering, leads us to a philosophically unsatisfactory situation: \textit{We will only see, what we expect to see. }
To overcome this limitation, our next steps should involve using generative models \cite{GAN} to create possible waveforms that could take us beyond our current expectations. 
This approach not only eliminates the need for a traditional template bank but also opens the possibility of generating new types of waveforms that have not yet been hypothesized. 
Detecting these waveforms, which might otherwise go unnoticed without specific candidate waveforms, opens the door to the new physics that could account for these observations. Paradoxically, by adopting modern AI techniques, we may return to an ’ancient’ way of doing science—where experiments precede theory, and discovery drives understanding.

\section*{Acknowledgments}
I would like to thank Matias Zaldarriaga and Aaron Zimmerman for their insightful comments. I am deeply grateful to Ira Rothstein for his invaluable discussions and encouragement, which had a significant impact on the development of this work.

\section*{Appendix}
For technical details and codes of model developent, training, noise generation, fine-tuning and inference see the shared Google Colab notebook. 
\url{https://colab.research.google.com/drive/1kKoxYlaQBuH61U8BNeCA0PTAwqMcI3uC}

%\section*{Codes}
%For technical details and codes of model developent, training, noise generation, fine-tuning and inference see the shared Google Colab notebook. 
%\url{https://colab.research.google.com/drive/1kKoxYlaQBuH61U8BNeCA0PTAwqMcI3uC}

\section*{Plots}

%%%%%%%%%%%%%%%%%%%%%%%%%%%%%%%%%%%%%%%%%%%%%%%%%%%%%%%%%%%%

\appendix
\newpage

%%%%%%%%%%%%%%%%%%%%%%%%%%%%%%%%%%%%%%%%%%%%%%%%%%%%%%%%%%%%

\end{document}